\documentclass[twocolumn,english,pre]{revtex4}
\usepackage[T1]{fontenc}
\usepackage[latin1]{inputenc}
\usepackage{graphicx}

\makeatletter

\usepackage{graphicx}
\usepackage{graphicx}

\usepackage{epsf}

\usepackage{babel}
\makeatother
\begin{document}

\title{Non-ergodic Intensity Correlation Functions for Blinking Nano Crystals }

\author{G. Margolin and E. Barkai}

\affiliation{Department of Chemistry and Biochemistry, Notre Dame University,
Notre Dame, IN 46556.}

\date{\today{}}

\begin{abstract}
We investigate the non-ergodic properties of blinking nano-crystals
using a stochastic approach. We calculate the distribution functions
of the time averaged intensity correlation function and show that
these distributions are not delta peaked on the ensemble average correlation
function values; instead they are W or U shaped. Beyond blinking nano-crystals
our results describe non-ergodicity in systems stochastically modeled
using the L\'{e}vy walk framework for anomalous diffusion, for example
certain types of chaotic dynamics, currents in ion-channel, and single
spin dynamics to name a few. 
\end{abstract}
\maketitle
Statistics of fluorescence intensity signals from single molecules,
atoms, and nanocrystals are in many cases analyzed using intensity
correlation functions (e.g., \cite{Verberk,GB1}). These correlation
functions are used to investigate a wide range of dynamical behaviors,
for example effects like anti-bunching and the stochastic dynamics
of large molecules coupled to their environment (\cite{Wolf,Annual}
and Ref. therein). In standard theories it is assumed that the process
of photon emission is stationary and ergodic. In contrast, measurements
of intensity correlation functions obtained from single nano-crystals
(NCs) exhibit a non-stationary and non-ergodic behavior \cite{Dahan,Brokmann};
as such these systems exhibit a statistical behavior very different
than other single emitting objects. 

More specifically, the fluorescence emission of single colloidal NCs,
e.g. CdSe quantum dots, exhibits interesting intermittency behavior
\cite{Nirmal}. A standard method to analyze such intensity signals
is to define a threshold $I_{th}$ and define two states: \emph{on}
if $I(t)>I_{th}$ and \emph{off} otherwise. For capped NCs (e.g. CdSe(ZnS)
core-shell NC) \emph{on} and \emph{off} times exhibit power law statistics
\cite{Kuno,Ken}, their probability density function (PDF) behave
like $\psi(\tau)\propto\tau^{-(1+\theta)}$ for large $\tau$, and
$\theta<1$. For example in \cite{Brokmann} $215$ NCs were measured
and the exponents $\theta_{on}=0.58\pm0.17$ and $\theta_{off}=0.48\pm0.15$
were found (note that within error of measurement $\theta_{on}=\theta_{off}=\theta$),
further all NCs are reported to be statistically identical. Since
$\theta<1$ the average \emph{on} and \emph{off}  times are infinite.
The divergence of occupation times naturally leads to non-ergodicity
in the blinking NCs \cite{Dahan,Brokmann} (see also \cite{Bouchaud}
for related discussion). Other measurements classify the intermittency
based on time average correlation function or closely related power
spectrum \cite{Dahan,Oijen,Pelton04}. In this type of analysis of
the experimental data there is no need to introduce a threshold value
$I_{th}$. 

From a single realization of intensity trajectory $I(t)$, recorded
in a time interval $(0,T')$, we may construct the time averaged (TA)
correlation function \begin{equation}
C_{TA}(t',T')=\frac{\int_{0}^{T}I(t+t')I(t)\textrm{d}t}{T},\label{eq01}\end{equation}
 where we denoted $T=T'-t'$. On the other hand we may generate many
intensity trajectories one at a time, and then average to obtain the
ensemble average correlation function $\langle I(t)I(t+t')\rangle$.
Single molecule experiments investigate time average correlation functions.
If the random process $I(t)$ is ergodic, the time average and the
ensemble average correlation functions are identical in statistical
sense, provided that the measurement time is long. Theories of correlation
functions of single emitting objects are many times based on the assumption
that the single molecule intensity trajectories are ergodic and hence
for simplicity, theories concentrate on the calculation of the ensemble
average correlation function. In this Letter we quantify the non-ergodic
properties of the correlation functions of blinking NCs, using a stochastic
approach. A related question of non-stastionarity, or more specifically
aging of the ensemble average correlation function, is a subject of
intensive research in the literature \cite{GB1,GL,Bald,LineShape04}.

\begin{figure}
\begin{center}\includegraphics[%
  width=1.0\columnwidth,
  height=0.65\columnwidth]{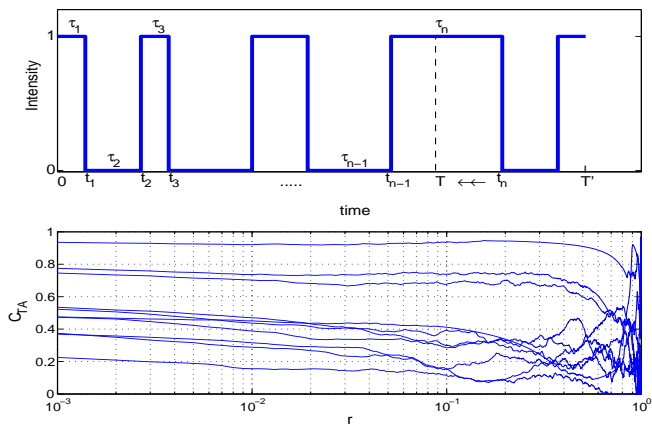}\end{center}

\caption{(a) A schematic representation of intensity blinking, note that we
redefine $t_{n}$ to be equal to $T$. (b) Ten numerically generated
realizations of the correlation function $C_{TA}(t',T')$ versus $r=t'/T'$,
for $\theta=0.8$. The correlation functions exhibit non-ergodic behavior
and are random, for ergodic processes all the ten time averaged correlation
functions would follow the same master curve, namely the ensemble
average correlation function. }

\label{fig1}
\end{figure}

We use a simple two state stochastic model, with which correlation
functions and non-ergodicity of the NCs are investigated. The intensity
$I(t)$ jumps between two states $I(t)=1$ and $I(t)=0$. At start
of the measurement $t=0$ the NC begins in state \emph{on} $I(0)=1$.
A schematic realization of the intensity fluctuations is shown in
Fig. \ref{fig1}. The process is characterized based on the sequence
$\{\tau_{1}^{on},\tau_{2}^{off},\tau_{3}^{on},\tau_{4}^{off},\cdots\}$
of \emph{on} and \emph{off} sojourn times or equivalently according
to the dots on the time axis ${t_{1},t_{2},\cdots}$, on which transitions
from \emph{on} to \emph{off} or vice versa occur (See Fig. \ref{fig1}).
The sojourn time $\tau_{i}$ is an \emph{off} time if $i$ is even,
it is an \emph{on} time if $i$ is odd. The times $\tau_{i}$ are
drawn at random from the PDF $\psi(\tau)$. These sojourn times are
mutually independent, identically distributed random variables. We
will consider the case $\psi(\tau)\propto\tau^{-(1+\theta)}$ where
$0<\theta<1$. When $\theta>1$, or when $\psi(\tau)$ is exponential,
we find an ergodic behavior. Note that similar intermittency behavior
describes a wide range of physical systems and models. In particular
the model describes a L\'{e}vy walk, which is an important stochastic
model for anomalous diffusion describing dynamics of systems mentioned
in the abstract \cite{GB1,Jung,Schlesinger}. The power law behavior
of the sojourn times of the NCs can be explained based on simple physical
models. For example random energy trapping models and random walks
models \cite{Remark} can be used to model such behavior, though currently
it is not clear yet what is the physical mechanism behind the observed
power law behavior \cite{GB1,Ken,Oijen}. 

In Fig. \ref{fig1}, ten typical simulated correlation functions are
plotted, the most striking feature of the figure is that the correlation
functions are random. These correlation functions are similar to those
obtained in the experiment \cite{Dahan}. Mathematically, the question
of non-ergodicity may be formulated in the following way. Since the
process $I(t)$ is random the time average correlation function $C_{TA}(t',T')$
is random. For ergodic processes, and in the long measurement time
limit, the distribution of $C_{TA}(t',T')$ is delta peaked and centered
around the ensemble average correlation function. For non-ergodic
processes the goal is to obtain the non-trivial limiting distributions
of $C_{TA}(t',T')$ which differ from the narrowly peaked delta functions
found for ergodic processes. In what follows we will denote $P_{C_{TA}(t',T')}(x)$
to be the PDF of $C_{TA}(t',T')$. 

To start our analysis we rewrite the time average correlation function
as \begin{equation}
C_{TA}(t',T')={\frac{\sum_{i\,\,\mbox{odd}}^{n}\int_{t_{i-1}}^{t_{i}}I(t+t'){\textrm{d}}t}{T}},\label{eq03}\end{equation}
 where we used the initial condition that $I(t)=1$ at time $t=0$.
Hence $I(t)=1$ in $t_{i-1}<t<t_{i}$ when $i$ is odd, otherwise
it is zero. The summation in Eq. (\ref{eq03}) is over odd $i$'s,
and $t_{n}=T$, namely $n-1$ in Eq. (\ref{eq03}) is the random number
of transitions in the interval $[0,T]$. From Eq. (\ref{eq03}) we
see that the time averaged correlation function is a sum of the random
variables \begin{equation}
\int_{t_{i-1}}^{t_{i}}I(t+t'){\textrm{d}}t=\left\{ \begin{array}{ll}
\tau_{i}-t'+{\mathcal{I}}_{i}t' & i\,\,\mbox{odd}\,\, t_{i}-t_{i-1}>t'\\
{\mathcal{I}}_{i}\tau_{i} & i\,\,\mbox{odd}\,\, t_{i}-t_{i-1}<t'\\
0 & i\,\,\mbox{even}\end{array}\right.\label{eq04}\end{equation}
 where\begin{equation}
{\mathcal{I}}_{i}=\left\{ \begin{array}{cc}
{\frac{\int_{t_{i}}^{t_{i}+t'}I(t){\textrm{d}}t}{t'_{\,}}} & \mbox{if}\,\, t_{i}-t_{i-1}>t'\\
{\frac{\int_{t_{i-1}+t'}^{t_{i}+t'}I(t){\textrm{d}}t}{\tau_{i}}} & \mbox{if}\,\, t_{i}-t_{i-1}<t'.\end{array}\right.\label{eq05}\end{equation}
 The ${\mathcal{I}}_{i}$'s are time averages of the signal $I(t)$
over periods of length $t'$ or $\tau_{i}=t_{i}-t_{i-1}$. Using Eqs.
(\ref{eq03},\ref{eq04}) we find an exact expression for the correlation
functions in terms of $\{\tau_{i}\}$ and $\{{\mathcal{I}}_{i}\}$\begin{equation}
TC_{TA}\left(t',T'\right)=\sum_{i\,\,\mbox{odd}}^{n}\tau_{i}-\sum_{\begin{array}{c}
i\,\,\mbox{odd}\\
\tau_{i}<t'\end{array}}^{n}(1-{\mathcal{I}}_{i})\tau_{i}-t'\sum_{\begin{array}{c}
i\,\,\mbox{odd}\\
\tau_{i}>t'\end{array}}^{n}\left(1-{\mathcal{I}}_{i}\right).\label{eq06}\end{equation}
 The first term on the right hand side of this equation is $T^{+}$
the total time spent in state \emph{on} in the time interval $[0,T]$,
in the remaining two terms we have considered sojourn times $\tau_{i}$
larger or smaller than $t'$ separately. 

\begin{figure}
\begin{center}\includegraphics[%
  width=1.0\columnwidth,
  height=0.65\columnwidth]{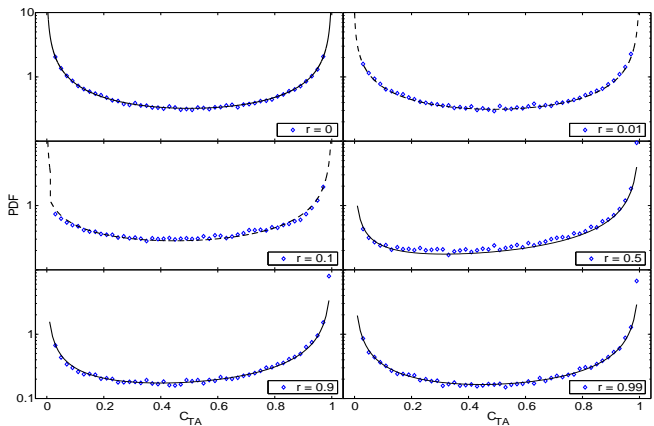}\end{center}

\caption{The PDF of $C_{TA}(t',T')$ for $\theta=0.3$ and different values
of $r=t'/T'$. The diamonds are numerical simulations and the curves
are analytical expression obtained for: (a) $r=0$, Eq. (\protect{}\ref{eq08})
solid curve, (b) $r=0.01,0.1$, Eq. (\protect{}\ref{eq12}) dashed
curve, and (c) for $r=0.5,0.9,0.99$, Eq. (\protect{}\ref{eq17})
solid curve. In the ergodic phase the PDF of $C_{TA}(t',T')$ would
be peaked around the ensemble average correlation function, which
for $r=0$ falls on $1/2$ and for $t'\rightarrow\infty$ is on $1/4$
(for any $r\neq0$). We see that any measurement is highly unlikely
to yield the ensemble average when $\theta=0.3$. }

\label{fig2}
\end{figure}

\begin{figure}
\begin{center}\includegraphics[%
  width=1.0\columnwidth,
  height=0.65\columnwidth]{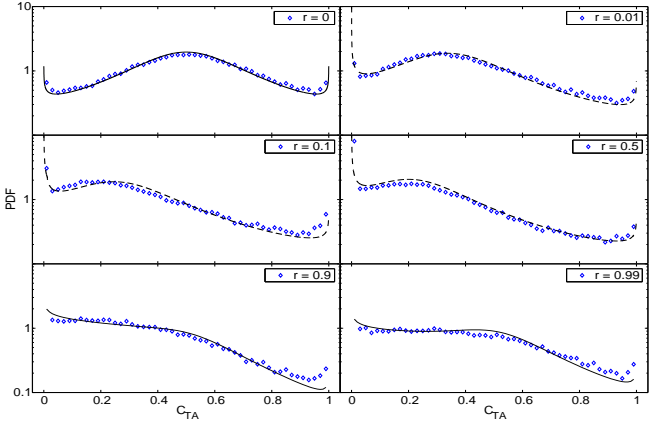}\end{center}

\caption{Same as Fig. \protect{}\ref{fig2} however now $\theta=0.8$. If
compared with the case $\theta=0.3$, the distribution function exhibits
a weaker non-ergodic behavior, namely for $r=0$ the distribution
function peaks on the ensemble average value of $1/2$. }

\label{fig3}
\end{figure}

We now illustrate the rich behaviors of the PDF $P_{C_{TA}(t',T')}(x)$
using numerical simulations, and later we consider the problem analytically.
We generate random realization of the process using $\psi(\tau)=\theta\tau^{-1-\theta}$
for $\tau>1$ and show two cases: $\theta=0.3$ in Fig. \ref{fig2}
and $\theta=0.8$ in Fig. \ref{fig3}. In both figures we vary $r\equiv t'/T'$.
The diamonds are numerical results which agree very well with the
theoretical curves, without any fitting. First consider the case $r=0$
in Figs. \ref{fig2},\ref{fig3}. For $\theta=0.3$ and $r=0$ we
see from Fig. \ref{fig2} that the PDF $P_{C_{TA}(t',T')}(z)$ has
a $U$ shape. This is a strong non-ergodic behavior, since the PDF
does not peak on the ensemble averaged value of the correlation function
which is $1/2$ for this case. On the other hand, when $\theta=0.8$
the PDF $P_{C_{TA}(t',T')}(z)$ has a $W$ shape, a weak non ergodic
behavior. To understand the origin of this type of transition note
that as $\theta\rightarrow0$ we expect the process to be in an \emph{on}
state or an \emph{off} state for the whole duration of the measurement,
hence in that case the PDF of the correlation function will peak on
$C_{TA}(t',T')=1$ and $C_{TA}(t',T')=0$ (i.e $U$ shape behavior).
On the other hand when $\theta\rightarrow1$ we expect a more ergodic
behavior, since for $\theta=1$ the mean \emph{on} and \emph{off}
periods are finite, this manifests itself in a peak of the distribution
function of $C_{TA}(t',T')$ on the ensemble average value of $1/2$
and a $W$ shape PDF emerges. Note that for $\theta<1$ there is still
statistical weight for trajectories which are \emph{on} or \emph{off}
for periods which are of the order of the measurement time $T'$,
hence the distribution of $C_{TA}(0,T')$ attains its maximum on $C_{TA}(0,T')=1$
and $C_{TA}(0,T')=0$. For $r>0$ we observe in both figures a non-symmetrical
shape of the PDF of the correlation function, which will be explained
later.

We first consider the non-ergodic properties of the correlation function
for the case $t'=0$. It is useful to define \begin{equation}
\mathcal{I}_{[a,b]}=\int_{a}^{b}I(t)\textrm{d}t/(b-a),\label{eq:Iab}\end{equation}
the time average intensity between time $a$ and time $b>a$. Using
Eq. (\ref{eq06}) and for $t'=0$ the time averaged correlation function
is identical to the time average intensity \begin{equation}
C_{TA}(0,T)={\mathcal{I}}_{[0,T]}={\frac{T^{+}}{T}}.\label{eq07}\end{equation}
 The random correlation function $C_{TA}(0,T)$ has a known asymptotic
behavior in the limit $T\rightarrow\infty$, found originally by Lamperti
\cite{Lamp} (see also \cite{GL}), this PDF is denoted with $\lim_{T\rightarrow\infty}P_{C_{TA}(0,T)}(x)=l_{\theta}(x)$,
and \begin{equation}
l_{\theta}\left(x\right)={\frac{\sin\pi\theta}{\pi}}{\frac{x^{\theta-1}\left(1-x\right)^{\theta-1}}{x^{2\theta}+(1-x)^{2\theta}+2x^{\theta}\left(1-x\right)^{\theta}\cos\pi\theta}},\label{eq08}\end{equation}
 for $0\leq x\leq1$. The transition between the $U$ shape behavior
and the $W$ shape behavior happens at $\theta_{c}=0.5946...$. The
Lamperti PDF is shown in Figs. \ref{fig2} and \ref{fig3} for the
case $r=0$, together with the numerical results. 

We now consider an analytical approach for the case $t'\ll T$. The
behavior of $P_{C_{TA}(t',T)}(x)$ for $t'\ne0$ is non-trivial, because
the ${\mathcal{I}}_{i}$'s in Eq. (\ref{eq06}) depend statistically
on the random variables ${\tau_{i}}$. To treat the problem we use
a non-ergodic mean field approximation. We noticed already that ${\mathcal{I}}_{i}$
defined in Eq. (\ref{eq05}) are short time averages of the intensity,
hence using mean field theory approach we replace the ${\mathcal{I}}_{i}$
in Eq. (\ref{eq06}) with the \emph{time} average intensity ${\mathcal{I}}_{[0,T]}$,
\emph{specific for a given realization}. Replacing ${\mathcal{I}}_{i}$
with the \emph{ensemble} average intensity is not appropriate. Hence
within mean field \begin{equation}
TC_{TA}(t',T')={\mathcal{I}}_{[0,T]}T-\left(1-{\mathcal{I}}_{[0,T]}\right)(t'N^{+}+\sum_{\begin{array}{c}
i\,\,\mbox{odd}\\
\tau_{i}<t'\end{array}}^{n}\tau_{i})\label{eq09}\end{equation}
 where $N^{+}$ is number of odd (i.e. \emph{on}) intervals satisfying
$\tau_{i}>t'$ and $i\leq n$. 

We now investigate the distribution of $C_{TA}(t',T')$ using the
approximation Eq. (\ref{eq09}), leaving certain details of our derivation
to a longer publication. First we replace $N^{+}$ with its scaling
form. Let $P(\tau>t')=\int_{t'}^{\infty}\psi(\tau){\textrm{d}}\tau$
be the probability of $\tau$ being larger than $t'$, we have $N^{+}\simeq KP(\tau>t')T^{+}/\int_{0}^{T^{+}}\tau\psi(\tau){\textrm{d }}\tau,$
where $K$ is a constant of order $1$, and $T^{+}/\int_{0}^{T^{+}}\tau\psi(\tau){\textrm{d }}\tau$
is total number of jumps in time interval $T^{+}$. A more refined
treatment yields \begin{equation}
N^{+}\simeq{\frac{\sin\pi\theta}{\pi\theta}}\left[\left({\frac{T^{+}}{t'}}\right)^{\theta}-1\right],\label{eq11}\end{equation}
 which is valid for $T^{+}/t'>1$. Similar scaling arguments are used
for the sum in Eq. (\ref{eq09}) which lead to \begin{equation}
\sum_{i\,\,\mbox{odd},\tau_{i}<t'}^{n}\tau_{i}\simeq\left(T^{+}\right)^{\theta}(t')^{1-\theta},\end{equation}
 an approximation which is valid for $t'<T^{+}$. For $t'>T^{+}$,
$N^{+}=0$ and $\sum_{i\,\,\mbox{odd},\tau_{i}<t'}^{n}\tau_{i}=T^{+}$.
In summary and after some rearrangements we obtain \begin{widetext}
\begin{equation}
 C_{TA} ( t' , T' ) \simeq 
\left\{
\begin{array}{l l}
 {\mathcal{I}}_{[0,T]} \left\{ 1  - \left(1 - {\mathcal{I}}_{[0,T]} \right)
\left[ \left( {r \over {(1-r)\mathcal{I}}_{[0,T]} } \right)^{1 - \theta} \left( { \sin \pi \theta \over \pi \theta} + 1 \right) - { \sin \pi \theta \over \pi \theta}{ r \over {(1-r)\mathcal{I}}_{[0,T]} } \right] \right\} 
&\ t' < T^{+} \\
{\mathcal{I}}_{[0,T]} ^2 &\ t' > T^{+}.
\end{array}
\right.
\label{eq12}
\end{equation}
\end{widetext} Eq. (\ref{eq12}) yields the correlation function, however unlike
standard ergodic theories the correlation function here is a random
function since it depends on ${\mathcal{I}}_{[0,T]}$. The distribution
of $C_{TA}(t',T')$ is now easy to find using the chain rule, and
Eqs. (\ref{eq07},\ref{eq08}, \ref{eq12}). In Figs. \ref{fig2},\ref{fig3}
we plot the PDF of $C_{TA}(t',T')$ (dashed curves) together with
numerical simulations (diamonds) and find excellent agreement between
theory and simulation, for the cases where our approximations are
expected to hold $r<1/2$. We observe that unlike the $r=0$ case
the PDF of the correlation function exhibit a non-symmetrical shape.
To understand this note that trajectories with short but finite total
time in state \emph{on} ($T^{+}\ll T$) will have finite correlation
functions when $t'=0$. However when $t'$ is increased the corresponding
correlation functions will typically decay very fast to zero. On the
other hand, correlation functions of trajectories with $T^{+}\sim T$
don't change much when $t'$ is increased (as long as $t'\ll T^{+}$).
This leads to the gradual nonuniform shift to the left, and {}``absorption''
into $C_{TA}(t',T')=0$, of the Lamperti distribution shape, and hence
to non-symmetrical shape of the PDFs of correlation function, in general,
whenever $r\ne0$. 

Finally, we turn to the case $T\ll t'$. Since $t'$ is large we use
a decoupling approximation and write Eq. (\ref{eq01}) as \begin{equation}
C_{TA}(t',T')\simeq{\mathcal{I}}_{[0,T]}{\mathcal{I}}_{[t',T']}.\label{eq13}\end{equation}
 We distinguish between two types of trajectories, those in which
no transition event occurs in the time interval $[T,T']$ and all
other trajectories. Let $P_{0}(a,b)$ be the probability of making
no transition between time $a$ and time $b$, also called the persistence
probability \cite{GL},\begin{equation}
P_{0}(a,b)\sim{\frac{\sin\pi\theta}{\pi}}\int_{0}^{a/b}x^{\theta-1}\left(1-x\right)^{-\theta}{\textrm{d}}x\label{eq14}\end{equation}
in the scaling limit. Using the Lamperti distribution for $\mathcal{I}_{[0,T]}$,
and probabilistic arguments with details left to the Appendix, we
find the PDF of $C_{TA}(t',T')$ \begin{widetext}
\begin{equation}
 P_{ C_{TA}(t',T') } \left( z \right)  \simeq 
\left[ 1 - P_0\left(T, T' \right) \right] 
\left\{ \left[ 1 - P_0\left( t' , T' \right) \right] \int_z^1 { l_{\theta} \left( x \right) \over x} {\textrm{d}} x + 
{ P_0\left( t',T' \right) \over 2} \left[ l_{\theta} \left( z \right) + \delta\left( z \right) \right] \right\}
+ P_0\left( T, T' \right) \left[ z l_{\theta}\left( z \right) + { \delta\left( z \right) \over 2} \right].
\label{eq17}
\end{equation}
\end{widetext} Note that to derive Eq. (\ref{eq17}) we used the fact that $\mathcal{I}_{[0,T]}$
and $\mathcal{I}_{[t',T']}$ are correlated. In Figs. \ref{fig2},\ref{fig3}
we plot these PDFs of $C_{TA}(t',T')$ (solid curves) together with
numerical simulations (diamonds) and find good agreement between theory
and simulation, for the cases where these approximations are expected
to hold, $r>1/2$. In the limit $t'/T'\rightarrow1$ Eq. (\ref{eq17})
simplifies to \begin{equation}
P_{C_{TA}(t',T')}(z)\sim[\ell_{\theta}(z)+\delta(z)]/2,\label{eq:halflamperti}\end{equation}
a result which is easily understood if one realizes that in this limit
$\mathcal{I}_{[t',T']}$ in Eq. (\ref{eq13}) is either 0 or 1 with
probabilities 1/2, and that the PDF of $\mathcal{I}_{[0,T]}$ is Lamperti's
PDF Eq. (\ref{eq08}).

To summarize, our work classifies the nonergodic properties of photoemission
intensity signal from NCs, and more generally L\'{e}vy walks, and
yields an analytical tool for the investigation of the nonergodic
correlation functions.

This work was supported by National Science Foundation award CHE-0344930.

~

\appendix

\section{Derivation of Eq.(15)}

To calculate the PDF of $C_{TA}(t',T')$ in Eq.(13) we use two steps: (i) calculate the PDF of \({\mathcal{I}}_{[t',T']}\) which statistically depends on \({\mathcal{I}}_{[0,T]}\) and then (ii) using the distribution of \({\mathcal{I}}_{[0,T]}\), which is the Lamperti's PDF Eq.(8), calculate the PDF of $C_{TA}(t',T')$.

Using the persistence probability, we approximate the conditional PDF of \({\mathcal{I}}_{[t',T']}\) for a given \({\mathcal{I}}_{[0,T]}\) in the case $T \ll t'$ by

\begin{widetext}

\begin{equation}
f_{ {\mathcal{I}}_{[t',T']} } (z|{\mathcal{I}}_{[0,T]}) \simeq 
\left[ 1 - P_0\left( T, T' \right) \right] 
Q_{ {\mathcal{I}}_{[t',T']} } 
\left(z\right) + P_0\left( T, T' \right) \left[
{\mathcal{I}}_{[0,T]}\delta\left( z - 1 \right) + \left( 1 -{\mathcal{I}}_{[0,T]}\right) ]\delta\left( z \right) \right],  
\label{eq15}
\end{equation}


where \(Q_{ {\mathcal{I}}_{[t',T']} } \left(z\right)\)
is the PDF of
\({\mathcal{I}}_{[t',T']}\)
conditioned that at least one transition occurs in
\([T,T']\).    In Eq. (\ref{eq15})
we introduced the correlation between 
\({\mathcal{I}}_{[t',T']}\)  and
\({\mathcal{I}}_{[0,T]}\) through the dependence 
of the right hand side of the equation on
\({\mathcal{I}}_{[0,T]}\). We assumed that in the case of no transitions in the time interval \([T,T']\), the probability of the interval \([t',T']\) to be all the time either $on$ or $off$ (the only possible choices) is linearly proportional to the value of \({\mathcal{I}}_{[0,T]}\).

The persistence probability controls also the behavior of


\begin{equation}
 Q_{ {\mathcal{I}}_{[t',T']} } \left(z\right) \simeq
\left[ 1 - P_0 \left( t' , T' \right) \right] \Theta\left(
0<z< 1\right) 
+ P_0\left( t', T' \right) {\delta\left( z \right) + \delta \left( z - 1 \right) \over 2}.
\label{eq16}
\end{equation}

\end{widetext}

Briefly, we assumed that if a transition occurs in the interval \([t',T']\) the distribution of \({\mathcal{I}}_{[t',T']}\)
is uniform [i.e., \(\Theta\left(
0<z< 1\right) =1\) if the condition in the parenthesis is correct]. This is a crude approximation which is, however, reasonable for our purposes (however when $\theta$ approaches 1, this approximation does not work).
The delta functions in Eq.(\ref{eq16}) arise from two types of trajectories: If no transition occurs either \({\mathcal{I}}_{[t',T']}=1\) (state $on$) or \({\mathcal{I}}_{[t',T']}=0\) (state $off$) with equal probability.

Based on Eq.(13), the PDF of $C_{TA}(t',T')$ is

\begin{equation} P_{C_{TA}(t',T')}(z)\approx\int_{0}^{1}\ell_{\theta}(x)f_{{\mathcal{I}}_{[t',T']}}\left(\left.\frac{z}{x}\right|x\right)\frac{dx}{x},\label{eq:PDF-Cst}\end{equation}
where we use the observation that \({\mathcal{I}}_{[0,T]}\) is distributed according to Lamperti distribution Eq.(8). Finally, from Eqs. (\ref{eq15},\ref{eq16},\ref{eq:PDF-Cst}) we obtain Eq.(15).

\end{document}